\documentclass[prl,amsmath,amssymb,showpacs,preprint]{revtex4}
\usepackage{graphicx}
\def\be{\begin{equation}}
\def\ee{\end{equation}}
\begin{document}
\title{Confinement into a state with persistent current by thermal quenching of loop of Josephson junctions}
\author{Jorge Berger}
\affiliation{Physics Department, Ort Braude College, P. O. Box 78,
21982 Karmiel, Israel} 
\email{phr76jb@tx.technion.ac.il}
\begin{abstract}
We study a loop of Josephson junctions that is quenched through its critical temperature. For three or more junctions, symmetry breaking states can be achieved without thermal activation, in spite of the fact that the relaxation time is practically constant when the critical temperature is approached from above. The probability for these states decreases with quenching time, but the dependence is not allometric. For large number of junctions, cooling does not have to be fast. For this case, we evaluate the standard deviation of the induced flux. Our results are consistent with the available experimental data.
\end{abstract}
\pacs{74.40.+k, 74.81.Fa, 05.70.Fh, 11.15.Ex}
\maketitle
We consider a process in which a superconducting loop that contains $n$ identical Josephson junctions is cooled through its critical temperature, in the absence of applied fields, and monitor the spontaneous generation of metastable states with persistent current. From the theoretical point of view, this process is enlightening, because it provides and additional example of a phase transition that is dominated by the time evolution of the system parameters, rather than by thermal equilibrium; this is a subject that is still far from being closed, and is thought to be relevant both to condensed matter physics and to cosmology. From the practical point of view, this process has significant importance, since flux trapping is a major obstacle for reproducible functioning of large scale ultra-high-speed superconductivity digital applications \cite{APL} and we would like to comprehend how it depends on the system parameters.

The best known theory for the description of dynamic phase transitions is the Kibble--Zurek mechanism (KZM) \cite{K,Z1,Z2}. It states that in these transitions the equilibrium critical scalings predict various aspects of the nonequilibrium dynamics of symmetry breaking, including the density of residual topological defects. 
Several numeric simulations have tested the predictions of the KZM, particularly the dependence of the density of defects on the quenching time. Some of these simulations lead to refinements of the KZM \cite{ref} and others disagree with it \cite{disag}. The KZM has also been tested in several experiments; among them we will mainly be interested on those performed in superconducting loops \cite{Carmi,Tafuri,Monaco}.

The study of a loop of Josephson junctions is appealing, because the ``rules of the game" are particularly simple \cite{tex1} and the system parameters can be tailored practically at will. This system may be neater and qualitatively different from other systems, because it can be trully divided into $n$ identical subsystems, whereas in other systems the division depends on a continuously varying coherence length. Moreover, experimental results are already avilable \cite{Carmi}.

The supercurrent through Josephson junction $i$ is given by
\be
I_{Ji}=I_c(t)\sin\gamma_i \;,
\label{dc}
\ee
where $\gamma_i$ is the gauge-invariant phase difference across the junction. $I_c$ vanishes above the critical temperature and increases when the temperature is lowered. We will consider a quenching process in which $I_c(t)$ grows from 0 to $I_{c0}$. This supercurrent may be interpreted as arising from a potential energy term $-\sum_i I_c\cos\gamma_i$. This potential energy gives rise to metastable asymmetric states in which the system can be trapped.

There are two typical processes for the formation of topological defects in the KZM. In one case defects become confined when the order parameter becomes unable to follow the change of the parameters of the system. Another scenario is activation due to thermal fluctuations close to the Ginzburg temperature. We shall see that the present system does not fit in either of these cases; the relaxation time does not diverge at the critical temperature and no activation energy is required in order to enter a metastable state.  
More precisely, let $R$ be the resistance of each junction, $L$ the self-inductance of the loop, let us assume that the capacitance is small and the resistance of the loop itself (above the critical temperature $T_c$) is much smaller than that of the junctions. Then there are two relaxation times in the problem: one of them is $\hbar/2eRI_c$, which is infinite regardless of the temperature above $T_c$, and the other is $L/nR$, which remains constant and refers to a process that does not attempt to align the order parameters of neighboring segments into the same phase. In a sense, our problem is similar to that of decompression of He$^4$ from the $\lambda$-line \cite{He}.

Let $I_{c0}$ be the maximal superconducting current through the junctions at low temperature and let us take $R$, $I_{c0}$, $\hbar$ and $2e$ as units. Accordingly, the units of voltage, energy, inductance and time will be $RI_{c0}$, $2eRI_{c0}$, $\hbar/2eI_{c0}$ and $\hbar/2eRI_{c0}$.

The state of the loop will be described by the set of values $\{\gamma_i\}$. The sum of these values can be interpreted as minus the magnetic flux enclosed by the loop. Since we assume that there is no applied magnetic flux,
\be
\sum_{i=1}^n\gamma_i=-LI \;,
\label{flux}
\ee
where $I$ is the current around the loop. Our goal is to find the probabilities for the possible values of $\sum_{i=1}^n\gamma_i$ after the loop has been cooled.

The rules for the evolution of $\{\gamma_i\}$ are stated in several textbooks \cite{tex1}. The ac Josephson relation is
\be
d\gamma_i/dt=V_i \;,
\label{ac}
\ee
where $V_i$ is the voltage across junction $i$. The total current through junction $i$ is 
\be
I=I_{Ji}+V_i+CdV_i/dt+I_{Ni} \;,
\label{totcurr}
\ee
where $I_{Ji}$ is the supercurrent, given by Eq.~(\ref{dc}), $C$ is the capacitance of each junction, and $I_{Ni}$ is the Johnson current. 
We will assume that during a period of time $\tau_1$ the system is kept above the critical temperature and $I_c=0$, then, during a period of time $\tau_2$ the system is quenched and $I_c$ grows up to $I_{c0}$ and finally, during a period of time $\tau_3$, $I_c=I_{c0}$.

In most of our calculations we will assume that the capacitance is negligible. In this case,
from Eqs. (\ref{flux}), (\ref{ac}) and (\ref{totcurr}),
\be
\frac{d\gamma_i}{dt}=-I_{Ji}-I_{Ni}-\frac{1}{L}\sum_{i=1}^n\gamma_i \;.
\label{evol}
\ee
The case $C\neq 0$ will be discussed below. We integrate Eq.~(\ref{evol}) by Euler iterations. For this purpose we divide the process into short periods of time of duration $\Delta t$.
The Johnson current averaged over a single period is given by
\be
I_{Ni}=\eta g_i/\sqrt{\Delta t} \;,
\label{noise}
\ee
where $g_i$ is a random number with zero average and variance 1 and $\eta=(2k_BT/R)^{1/2}$, with $k_B$ Boltzmann's constant and $T$ the temperature. We assume that the temperature remains close to the critical temperature during the entire process; accordingly, $\eta$ will be taken as constant.

We first consider the case in which the loop is cooled instantaneously, i.e., $\tau_2=0$. If we ignore the Johnson current, Eq.~(\ref{evol}) is equivalent to viscosity-dominated motion of a particle in $n$-dimensional space that feels a potential energy $(1/2L)(\sum_{i=1}^n\gamma_i)^2-I_c\sum_{i=1}^n\cos\gamma_i$. In this situation, $\{\gamma_i\}$ evolves towards a local minimum of the potential energy. During the first stage, $I_c(t)=0$ and the only local minimum is the plane $\sum_{i=1}^n\gamma_i=0$. During the last stage, $I_c(t)=1$. At this stage the absolute minimum is located at the origin, $\gamma_i=0$, but for sufficiently large values of $L$ additional local minima may also exist. We ask whether the values $\{\gamma_i\}$ could wander in the plane $\sum_{i=1}^n\gamma_i=0$ and then, when the temperature is lowered, flow into a local minimum different from $\gamma_i=0$. For our present purpose, two states such that their respective values of $\gamma_i$ differ by integer multiples of $2\pi$ and $\sum_{i=1}^n\gamma_i$ is the same for both, will be considered equivalent.

\begin{figure}
\scalebox{0.85}{\includegraphics{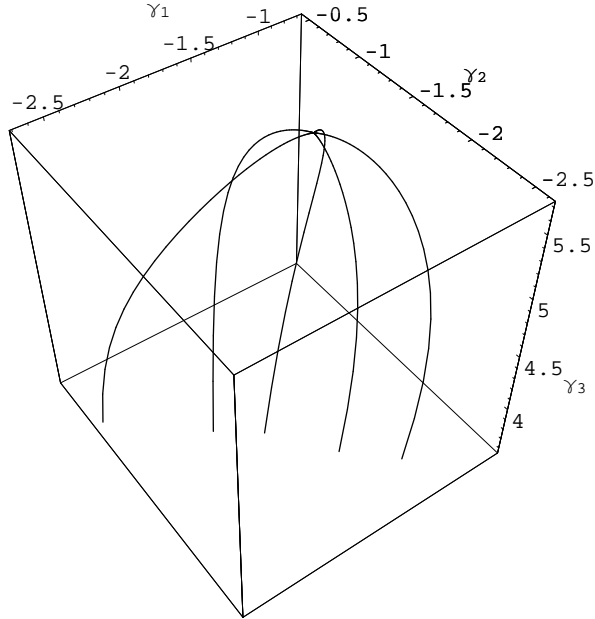}}%
\caption{\label{flow}Several evolution lines for the case of three junctions, for $L=4$ and in the absence of thermal fluctuations. All these curves start at the plane $\gamma_1+\gamma_2+\gamma_3=0$ and converge to the point $\gamma_1=\gamma_2=-0.98,\gamma_3=2\pi-0.98$.}
\end{figure}

Figure \ref{flow} shows that this situation is possible for the case $n=3$. The figure shows five evolution curves that start at the plane $\sum_{i=1}^n\gamma_i=0$ and, in the absence of thermal fluctuations, flow to a local minimum of the potential energy. Note that although the temperature is assumed to change instantaneously, $\sum_{i=1}^n\gamma_i$ builds up during a lapse of time of the order of $\hbar/eRI_{c0}$.

We studied the evolution including thermal fluctuations, for $n=3$ and $n=4$, for several values of $\eta$. In all cases, we started from values of $\{\gamma_i\}$ randomly located in the interval $-\pi <\gamma_i<\pi$.
In order to achieve initially an equilibrium distribution, evolution was followed during a period $\tau_1$ with $I_c=0$. After that, evolution was followed during a period $\tau_3$ with $I_c=I_{c0}$. In order to decide what is the ``final" state, we should average over an additional period of time, in order to filter out thermal fluctuations. We found it easier to turn off at this stage the fluctuations and let the state converge to the nearest local minimum. For each set of values, this process was reapeated 1000 times and the probability for confinement in a given state was evaluated as the number of times this state was obtained, divided by 1000. In most cases the final state was the ground state $\gamma_i=0$, but the first excited state was also reached. Due to the symmetry of the problem, these states are degenerate, i.e., the $\gamma_i$'s can be permuted and all the signs can be inverted. For the parameters we considered, we did not find cases with higher excited states.

\begin{figure}
\scalebox{0.85}{\includegraphics{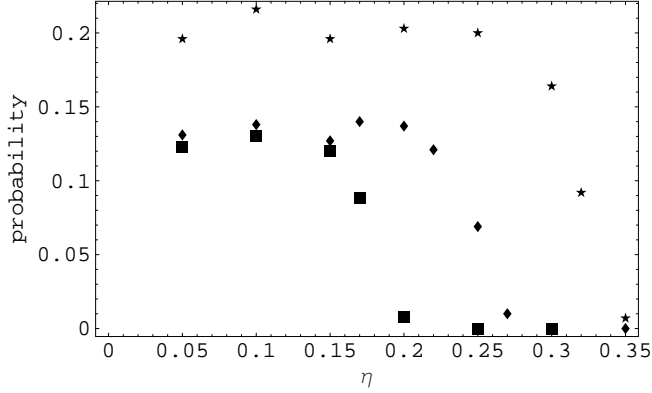}}%
\caption{\label{probs}Probability for a current-carrying metastable state as a function of the size of thermal fluctuations. Parameters used: $\tau_1=40000$, $\tau_2=0$, $\tau_3=10000$, $\Delta t=0.1$, $C=0$. $\Diamond$ $n=3$, $L=4$; $\star$ $n=4$, $L=4$; $\Box$ $n=4$, $L=2$.}
\end{figure}

The results are shown in Fig.~\ref{probs}. We avoided values of $\eta$ that might be too small to enable thermalization during the period $\tau_1$. The probabilities shown in the graph correspond to the total probability of reaching any of the (degenerate) excited states. As a general trend, we see that the probability of ending at an excited state increases with the number of junctions and with the normalized self-inductance. We also see that this probability is fairly independent of the size of thermal fluctuations, until a sufficiently large value of $\eta$ is reached. Beyond this value, there is a fast decrease of this probability.

It is reasonable to expect that the probability for the metastable state will decrease when the thermal energy $k_BT$ becomes comparable to the energy barrier that confines this state, i.e., the difference between the energy at the saddle-point and the energy at the local minimum. For $n=3$, $L=4$, a local minimum is at $\gamma_1=\gamma_2=-0.98,\gamma_3=2\pi-0.98$, the corresponding saddle point is at $\gamma_1=\gamma_2=-0.67,\gamma_3=\pi+0.67$ and the energy difference is 0.25; for $n=4$, $L=4$, a local minimum is at $\gamma_1=\gamma_2=\gamma_3=-0.83,\gamma_4=2\pi-0.83$, the saddle point is at $\gamma_1=\gamma_2=\gamma_3=-0.54,\gamma_4=\pi+0.54$ and the energy difference is 0.42; similarly, for $n=4$, $L=2$, the energy barrier is 0.13. In all cases we find that the probability for the metastable state decreases to about half its maximum value when the thermal energy is about one eighth of the barrier energy. Clearly, the precise value depends on $\tau_3$; in principle, for $\tau_3\rightarrow\infty$, the metastable state should always decay. If the thermal energy becomes of the order of the energy difference between the excited and the ground state, then the probability for the excited state will increase with temperature (equilibrium probability), but we are not interested in this regime.

\begin{figure}
\scalebox{0.85}{\includegraphics{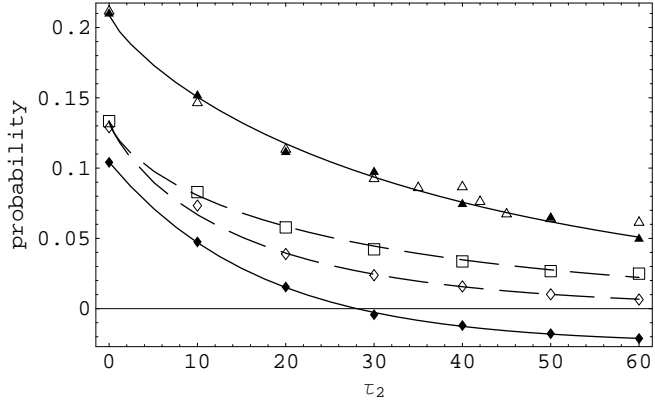}}%
\caption{\label{time4}Probability for a metastable state as a function of the cooling time for loops with 3 or 4 junctions. 
The empty (filled) symbols correspond to $I_c(t)$ proportional to $t$ (to $t^2$) and are fitted by dashed (continuous) lines. For visibility, the line for $n=3$, $L=4$ and $I_c(t)\propto t^2$ has been lowered by 0.025. Parameters used: $\tau_1=40000$, $\tau_2+\tau_3=10000$, $\Delta t=0.1$, $C=0$. Unless stated otherwise, $\eta =0.1$. $\Diamond$ $n=3$, $L=4$;  $\Box$ $n=4$, $L=2$; $\triangle$ $n=4$, $L=4$, $\eta =0.2$.}
\end{figure}

\begin{figure}
\scalebox{0.85}{\includegraphics{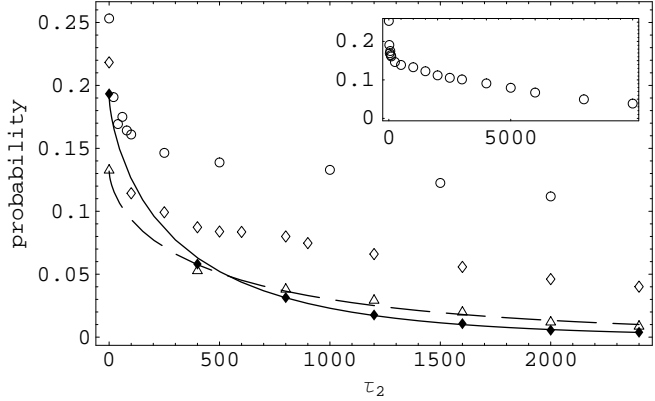}}%
\caption{\label{time6} Like Fig.~\ref{time4}, for $n=5$ and $n=6$. The inset shows $P(\tau_2)$ in the range $0\le\tau_2\le 10^4$. For visibility, the line for $n=5$, $L=2$ and $I_c(t)\propto t$ has been raised by 0.025.$\Diamond$ $n=5$, $L=2$; $\triangle$ $n=5$, $L=1$; $\circ$ $n=6$, $L=2$.}
\end{figure}

Let us now study the influence of the cooling time on the probability for a metastable state. At the moment that $I_c(t)$ becomes different from zero, there will be an incentive for leaving this state; on the other hand, as long as $I_c(t)$ is small, the confining barrier will also be small and the way out will be easy. We therefore expect that the trapping probability will decrease with $\tau_2$. In most of our calculations we assumed that $I_c(t)$ is proportional to the temperature below $T_c$ and therefore increases linearly with $t$, but we also considered the case $I_c(t)\propto t^2$, which is more realistic for strong coupling \cite{Monaco}. For the present purpose, simulations were repeated $10^4$ times. Figures \ref{time4} and \ref{time6} show our results for several values of $n$, $L$ and $\eta$. 

The topological charge for a given state may be defined as the sum of the topological charges of all junctions, where the topological charge of junction $i$ is the closest integer to $\gamma_i/2\pi$. For all the cases considered in Figs. \ref{time4}--\ref{time6}, the topological charge  was 0 or $\pm 1$.

The probabilities shown in Figs. \ref{time4}--\ref{time6} are also the expectations of the absolute value of the topological charge. It is therefore tempting to identify this probability with the density of defects, and anticipate that it will decrease as a power of $\tau_2$. However, the arguments that lead to the time dependence of the defect density in \cite{Z2} seem to be irrelevant in the present case; there is no obvious way to associate the presence of topological charge to some primordial coherence length and, indeed, our results cannot be fitted by a power dependence. Denoting the probability by $P$, most of the curves in our results (typically for small $n$ and $L$) can be fitted by the empiric form $P(\tau_2)\propto\exp[-(\tau_2/\tau_0)^{\sqrt{2/n}}]$ in the case $I_c(t)\propto t$ and by $P(\tau_2)\propto\exp[-(\tau_2/\tau_0)^{\sqrt{2/(n-1)}}]$ in the case $I_c(t)\propto t^2$. The characteristic time $\tau_0$ depends very strongly on the number of junctions and only weakly on the size of the energy barrier or on the temperature. For $n=3$, $16\alt\tau_0\alt 17$; for $n=4$, $26\alt\tau_0\alt 39$; for $n=5$, $350\alt\tau_0\alt 540$.

Our empirical fits suggest that for $n\gg 1$ the probabilities for metastable states decrease very slowly with the cooling time. Indeed, in the experiment that involved 214 junctions \cite{Carmi}, the distribution of permanent currents was found to be independent of the cooling time (up to the order of a minute).

Part of the probabilities shown in Figs. \ref{time4}--\ref{time6} do not decrease at a uniform rate. Instead, they seem to decay in two stages. A possible explanation might be that the region in phase space that in the absence of thermal fluctuations would flow into a metastable state can be divided into two subregions, such that escape from one subregion is much easier than escape from the other.

\begin{figure}
\scalebox{0.85}{\includegraphics{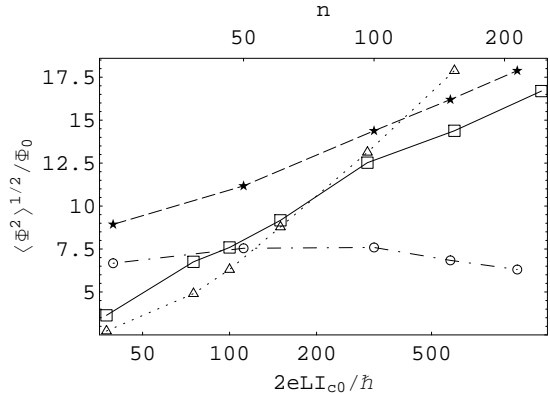}}%
\caption{\label{Large} Standard deviation of the induced flux, $\Phi /\Phi_0=-\sum_i\gamma_i$, as a function of the self-inductance and the number of junctions. Two curves are for fixed $n$, and $L$ is shown in the lower abscissa; the other two are for fixed $L$, and $n$ is shown in the upper abscissa. The abscissas are in logarithmic scale. The symbols have been joined for visibility. The parameters are as in Fig.~\ref{probs} and $\eta =0.1$. Each simulation was repeated 400 times. $\Box$ $n=100$; $\triangle$ $n=214$; $\circ$ $L=100$; $\star$ $L=600$.}
\end{figure}

Let us now consider large values of $n$ and $L$, as were encountered in the experiment. In this case many different final metastable states are possible, and the most significant experimental quantity will be the variance of the induced flux. Our results are shown in Fig.~\ref{Large}. As in the case of small values of $n$ and $L$, the general trend is increase of the typical flux with increase of either $n$ or $L$. However, these individual increases appear to saturate. For instance, for $L\alt 200$, the standard deviation of the flux actually decreases with $n$ in the range $100\alt n\alt 200$.

In the experiment \cite{Carmi}, $n=214$ and $\langle \Phi^2 \rangle^{1/2}/\Phi_0=7.4\pm 0.7$, where $\Phi$ is the induced flux and $\Phi_0$ the quantum of flux. Comparison with Fig.~\ref{Large} indicates that $2eI_{c0}L/\hbar$ should be in the range between $\sim 100$ and $\sim 150$. The estimates of Ref.~\cite{Carmi} suggest that $2eI_{c0}L/\hbar\sim 600$. Since the experimental estimate was not based on a measurement, but rather on a plausibility argument for the size of $I_{c0}$, and the junctions are not really all identical, the agreement is reasonable.

Let us finally consider the case $C\neq 0$. In this case we integrated Eqs. (\ref{ac}) and (\ref{totcurr}) as a system of differential equations. In digital applications, a preferred value is $C=\hbar /2eI_{c0}R^2$, which provides for fast switching without oscillations. We have repeated our calculations for this case and for several representative values of the other parameters. We found that a capacity of this size has no qualitative effect.

In summary, we have performed simulations that describe the formation of symmetry breaking states when a loop of Josephson junctions is quenched. Among the typical systems in which symmetry breaking occurs in a dynamics-dominated process, the present system constitutes a class of its own. Our results agree with the experiment in the case of large $n$ and raise predictions for the case of small $n$. 

This work has been supported by the Israel Science Foundation under grant 4/03-11.7. I am grateful to Alan Kadin for useful comments.

\end{document}